# Electronic Structure and Resonant X-ray Emission Spectra of Carbon Shells of Iron Nanoparticles

V. R. Galakhov[a], S. N. Shamin[a], E. M. Mironova[b], M. A. Uimin[a],
A. Ye. Yermakov[a], and D. W. Boukhvalov[c]

[a] Institute of Metal Physics, Ural Branch, Russian Academy of Sciences, Yekaterinburg, 620990 Russia
[b] Ural State Mining University, ul. Kuibysheva 30, Yekaterinburg, 620144 Russia
[c] Korea Institute for Advanced Study (KIAS), Hoegiro 87, Dongdaemun_Gu, Seoul 130-722, Republic of Korea

*The electronic structure of carbon shells of carbon encapsulated iron nanoparticles carbon encapsulated Fe@C has been studied by X-ray resonant emission and X-ray absorption spectroscopy. The recorded spectra have been compared to the density functional calculations of the electronic structure of graphene. It has been shown that an Fe@C carbon shell can be represented in the form of several graphene layers with Stone–Wales defects. The dispersion of energy bands of Fe@C has been examined using the measured C Kα resonant X-ray emission spectra.*

## INTRODUCTION

Metallic magnetic nanoparticles can be used in many engineering and biological fields: for drug delivery, cancer detection, catalysis, etc. Metallic nanoparticles are coated with carbon to prevent their oxidation and agglomeration under heat treatment.

Using X-ray absorption and photoelectron spectroscopy, we previously established that a core in Fe@C and Ni@C nanocomposites is in the metallic state [1], which, as was shown for Ni@C, was preserved for at least two years [2]. According to the ab initio calculations of the electronic structure and measurements of C 1*s* X_ray absorption spectra, the carbon shell in nanocomposites can be simulated by several graphene layers with Stone–Wales defects [2]. However, photoelectron spectroscopy and absorption spectroscopy involving the photoelectron yield mode are surface-sensitive methods. They allow analysis of only one nanoparticle layer in a compacted sample owing to a small mean free path of electrons. The surface layer of the sample is subjected to an external action; in particular, it interacts with functional groups and can differ from internal layers in composition and structure. In application to Me@C nanocomposites, X-ray emission (fluorescence) spectroscopy, which is a volume-sensitive method for studying the electronic structure of substances, has an advantage as compared to surface-sensitive photoelectron spectroscopy.

X-ray emission and absorption spectra are usually considered as independent and make it possible to obtain data on the electron energy structure of materials: according to selection rules, C $K\alpha$ X-ray emission and C $1s$ absorption spectra reflect the occupied and vacant $2p$ states of carbon, respectively. In the case of resonant inelastic X-ray scattering (RIXS), a coherent absorption–emission event occurs as a single-step process. Experimental RIXS spectra of carbon, which is a system with weak electronic correlations, can be described using the band model. When an inner electron is excited to the conduction band, X-ray emission occurs from the point of the Brillouin zone with the same quasi-momentum of the electron **k** in the valence band. The conservation of the momentum in RIXS processes determines the spectral features reflecting the dispersion of conduction and valence bands.

Resonant inelastic X-ray scattering spectra were measured for graphite in [3–6], for carbon nanotubes in [7–10], and for graphene on Pt in [11]. In [6, 12], RIXS spectra for SiC and graphite single crystals were used to obtain $E(\mathbf{k})$ dispersion dependences (band-mapping).

In this work, C $K\alpha$ resonant X-ray emission spectra are used for the first time to examine the electronic structure and to construct dispersion dependences $E(\mathbf{k})$ of the energy on the quasi-momentum for nanoparticles, in particular, Fe@C nanoparticles.

**FEATURES OF THE EXPERIMENT AND CALCULATION**

Iron nanoparticles capsulated into carbon were obtained by contactless levitation melting in a high-frequency field and by the evaporation of melted metal in the flow of an inert gas with hydrocarbons [1, 13, 14]. The size of resulting Fe@C particles was about 10 nm.

X-ray absorption and emission spectra were recorded on the I511-3 line of the MAX-lab II storage ring (Lund, Sweden). The C $1s$ absorption spectra were measured in the fluorescence yield mode. The electronic structure of Fe@C nanoparticles was calculated within the density functional theory using the QUANTUM_ESPRESSO code [15] and the GGA–PBE + vdW approximation [16], which is applicable for the description of the parameters of the crystal lattice of iron taking into account weak van der Waals forces [17] and makes it possible to correctly describe the atomic structure of fcc iron and its interaction with carbon [18]. The inclusion of weak interactions is necessary for the more accurate calculation of the metal–graphene distance [19]. Three graphene layers with defects were taken as the minimum model, which allows the description of the carbon coating of metallic nanoparticles. This model was tested in application to the Ni@C system [2]. The calculations for the graphene layer on the (111) surface of iron are compared to the results for the graphene layer with Stone–Wales defects.

**RESULTS AND DISCUSSION**

The C $K\alpha$ emission and C $1s$ absorption spectra of Fe@C are shown in Fig. 1 in comparison with the C $2p$ densities of states calculated for (*1*) the graphene layer on the (111) surface of iron, (*2*) the graphene layer with Stone–Wales defects, and (*3*) three graphene layers. Thus, we separated the following most important characteristics of capsulated particles (in the order of increasing importance): the contribution of the metal to the density of states of the carbon layer, the presence of Stone–Wales defects, and multilayer coating. We do not mention the contribution from contamination of the carbon layer by functional groups, which depend on the material storage time. Furthermore, the signals from functional groups located on the surface of the carbon layer are different for X-ray absorption spectra measured in the (surface-sensitive) photoelectron yield mode and (volume-sensitive) fluorescence mode, which is used in this work.

The C $K\alpha$ X-ray emission (fluorescent) spectrum was measured at excitation energy of about 300 eV, which corresponds to normal emission. Following the interpretation presented in [3, 20], the principal maximum and high-energy shoulder in the emission spectrum can be assigned to the $\sigma$ and $\pi$ orbitals, respectively. The C $1s$ absorption spectra of Fe@C are very similar to the spectra of graphite with pronounced $\pi^*$ and $\sigma^*$ bands, but with some minor differences. The SW feature marked by the arrow below the $\pi^*$ resonance is attributed to Stone–Wales defects [2]. It is absent in the spectrum of graphite and is observed in C $1s$ absorption spectra of defect graphene and carbon nanotubes [21–25]. The energy range of 287–289 eV can contain signals from –COOH and/or –CH groups contaminating the carbon shell of Fe@C [2, 26]. The contribution of signals from functional groups is marked in the figure as –OH. A signal from such contaminations was detected in C 1*s* X-ray absorption spectra of Ni@C nanoparticles measured in the total photoelectric effect yield mode two years after synthesis [2].

Although, when the density of states is calculated within the model implying the graphene layer on the (111) surface of iron, the number of metal atoms is equal to the number of carbon atoms and the number of layers is doubled, the contribution from iron to the electronic structure of the shell of Fe@C is very small. The contribution caused by the mixture of vacant Fe 3*d* and C 2*p* states is observed in the calculated density of states (line *1*) near the Fermi level (in the range from 0 to 1 eV). This contribution is invisible in the experimental spectrum. The energy gap between occupied and free C 2*p* states of Fe@C is absent. The calculated electron densities of states are in good agreement with the measured spectra of Fe@C. Resonant X-ray emission spectra will be analyzed below with the electron density of states calculated for three graphene layers.

Figure 2 shows C *K*α resonant X-ray spectra (RIXS) of Fe@C measured at various excitation energies, which were chosen according to the corresponding absorption spectra and are marked by arrows. The emission spectra exhibit an elastic peak, which exactly corresponds to the excitation energy, and an inelastic part, which is sometimes called Raman scattering. The probabilities of the elastic and inelastic scattering depend on the lifetimes of the main hole and excited electron. The probability of inelastic scattering is higher for electrons in delocalized states. Localized states in the conduction band are responsible for the elastic line. The energy of the high-energy maximum (the so-called elastic peak) is equal to the excitation energy. The spectra shown in Fig. 2 are normalized to the maximal intensity of the inelastic part.

It can be seen that C *K*α spectra of Fe@C are similar to the spectra of graphite, carbon nanotubes, and graphene [7, 11]. At excitation energies of 284–286 eV (spectra *a–c*), when electrons are excited to localized π states, the intensity of the elastic peak exceeds the intensity of the inelastic part of the spectrum. The intensity of the elastic peak decreases sharply at excitation energies sufficient for the excitation of a core electron to the σ band. A further increase in the excitation energy is accompanied by a smooth decrease in the intensity of the elastic peak. Spectrum *h* obtained after excitation by photons with an energy of 300 eV, which is much higher than the energy of the absorption edge, corresponds to normal emission and reflects the partial electron density of C $2p$ states that is integrated over the Brillouin zone.

To plot the dispersion curve $E(\mathbf{k})$ using RIXS spectra, it is necessary to separate the coherent part of the spectra formed by absorption–emission processes with the conservation of $\mathbf{k}$. We used the procedure proposed in [12]: the incoherent component of the spectra should be similar to the normal emission spectrum (spectrum $h$ obtained at excitation energy of 300 eV). The contribution of the incoherent part was taken to be the maximum possible under the condition of the positiveness of the difference spectrum in the entire energy range up to the region of the elastic line in the subtractive spectrum. Although the subtraction procedure is arbitrary, difference spectra allow the separation of peaks and features (specified by vertical lines in the figure), which were used to plot dispersion curves. The calculations of $E(\mathbf{k})$ for the defect graphene layers, as well as for graphene layers on the surface of iron, with the use of a supercell cannot be compared with the experimental data because of a high density of dispersion curves.

The dispersion curves calculated for three graphene layers are superposed with RIXS spectra as in Fig. 1. The vertical lines in Fig. 2 are drawn from elastic lines that correspond to excitation energies and are shown in the lower panel. Horizontal line segments (with the same vector $\mathbf{k}$) are drawn from the points of intersection of the vertical lines with dispersion curves and have lengths corresponding to the energy distances from the elastic lines to the features of the RICS spectra marked by short solid line segments in the lower panel of the figure.

The singular points thus obtained are shown on dispersion curves in the upper panel of Fig. 2. Different symbols correspond to spectra measured at different excitation energies. The experimental points closely correspond to the calculated dispersion curves except for the point at an energy of about –5eV, which is determined from spectrum $a$ and lies between dispersion curves. This point is likely attributed to the Stone–Wales defects, because spectrum $a$ was obtained when an electron was excited to the region of vacant states determined by defects (see, e.g., [2]). Stone–Wales defects are manifested as a small rise in the absorption spectrum, whereas the defect is manifested as a fairly intense peak in the emission spectrum *a*. Thus, RIXS spectra are more sensitive to the Stone–Wales defects than absorption spectra.

**CONCLUSIONS**

X-ray spectral studies and density functional calculations of the electronic structure have shown that the carbon shell of iron nanoparticles capsulated into carbon can be represented in the form of several graphene layers with Stone–Wales defects. The capabilities of RIXS spectroscopy have been demonstrated for the first time in application to the construction of the dispersion dependence $E(\mathbf{k})$ of the energy of electrons on the quasi-momentum of the carbon shells of iron nanoparticles capsulated into carbon. The presence of Stone–Wales defects is manifested both in the C 1*s* absorption spectrum and (more clearly) in the RIXS spectrum obtained at energy lower than the energy of the C 1*s* absorption edge of graphene.

We are grateful to Dr. Annette Pietzsch for assistance in the experiments on the I511-3 line of the MAX-lab II storage ring. D.W.B. acknowledges support from the Center for Advanced Computation, Korea Institute for Advanced Study. This work was supported in part by the Russian Foundation for Basic Research (project nos. 11-02-00166 and 10-02-00323).

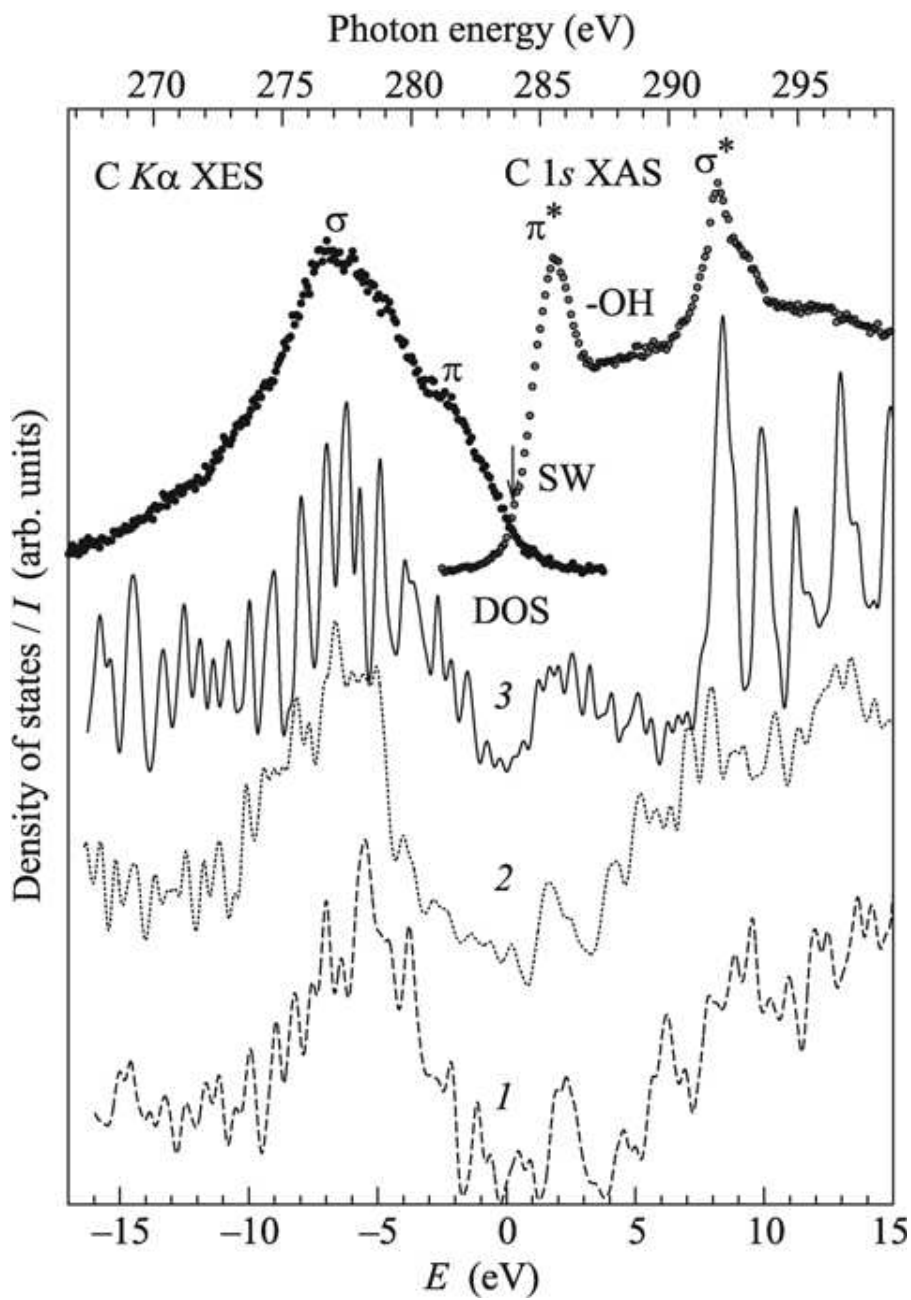

**Fig. 1.** C $K\alpha$ X-ray emission spectrum (XES, normal emission, excitation energy $h\nu_{exc}$ = 300 eV) and C 1s X-ray absorption spectrum (XAS) measured for Fe@C nanoparticles in comparison with the electron density of states calculated for (*1*) the graphene layer on the (111) surface of iron, (*2*) the graphene layer with Stone–Wales defects, and (*3*) three graphene layers. The experimental spectra and calculated density of states are matched in shape.

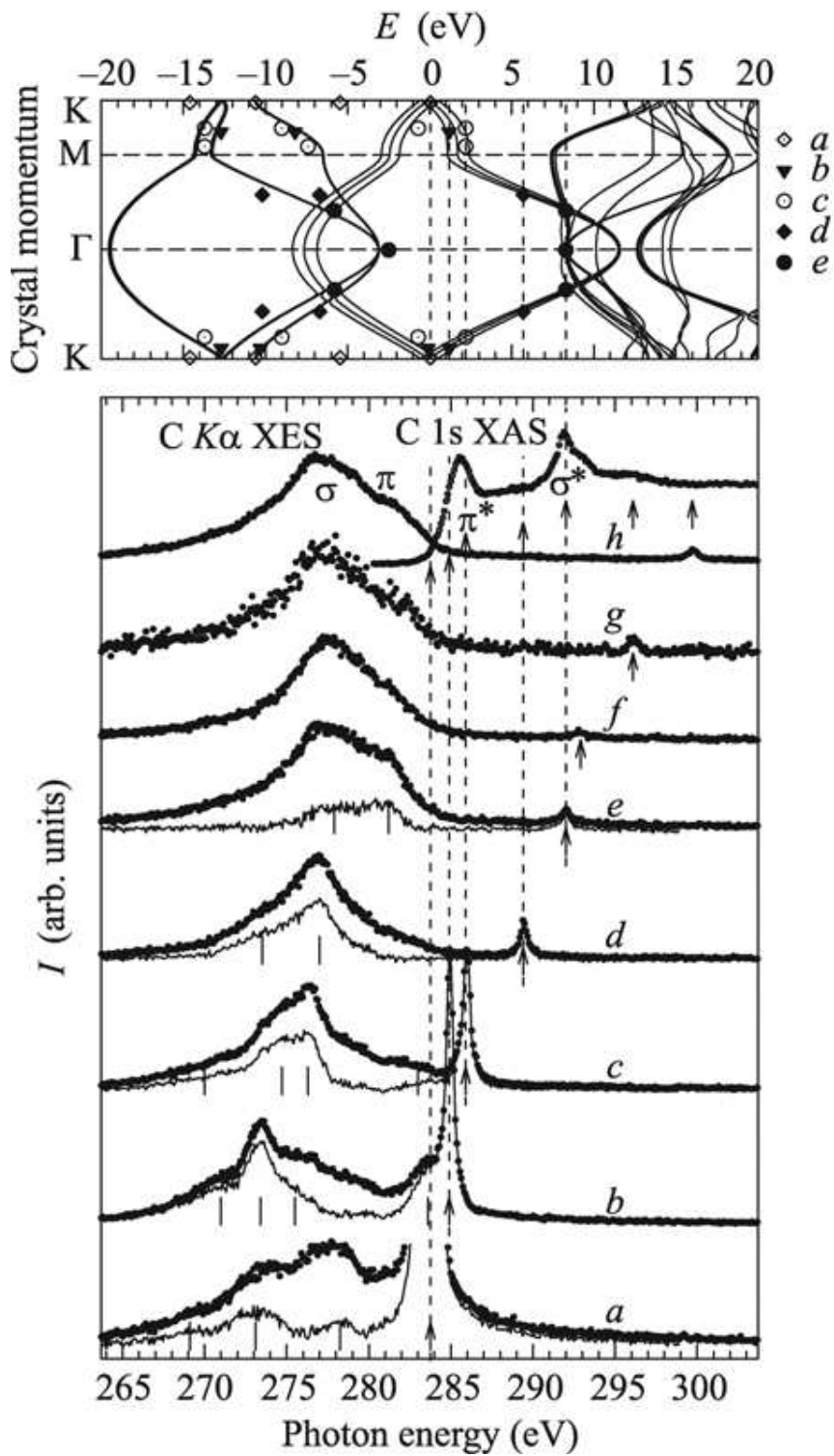

**Fig. 2.** (Lower panel) C 1s X-ray absorption spectrum (XAS) of Fe@C measured in the fluorescence yield mode, (points) C $K\alpha$ emission spectra of Fe@C measured at various excitation energies, and (solid lines) contributions from the coherent component to RIXS spectra. Arrows mark the excitation energies. The vertical straight lines specify the features of the spectrum that were used to plot dispersion curves. (Upper panel) Dispersion curves $E(\mathbf{k})$ calculated for three graphene layers with experimental points. Different symbols correspond to different excitation energies. The lower and upper scales refer to the experimental spectra and band calculation results, respectively.